\documentclass[a4paper,12pt]{article}
\usepackage[a4paper, total={6in, 10in}]{geometry}
\usepackage[table,xcdraw]{xcolor}
\usepackage{subcaption}
\usepackage{rotating}
\usepackage{amsfonts}
\usepackage[official]{eurosym}
\usepackage{stackrel}
\usepackage{caption}
\usepackage{graphicx}
\usepackage{xspace}
\usepackage{tikz}
\usepackage{mathtools}
\usepackage{url}
\usepackage{authblk}
\usepackage{hyperref}
\usepackage{amsthm}
\usepackage{bm}
\usepackage[title]{appendix}
\usepackage{bbold}
\usepackage{pdfpages}
\usepackage{graphicx}
\usepackage[english]{babel}
\usepackage{hyperref}
\usepackage{float}
\usepackage[sorting=none]{biblatex}
\usepackage{csquotes}

\mathtoolsset{showonlyrefs,showmanualtags}

\hypersetup{
	colorlinks=true,
	urlcolor=blue,
	citecolor=blue,
	linktoc=all,
	linkcolor=blue}
	
\definecolor{DarkGreen}{rgb}{0,0.4,0}
\definecolor{cobalt}{rgb}{0.0, 0.28, 0.67}

\newcommand{\rev}{``}

\title{A Bayesian algorithm for sample selection bias correction}
\author{Valerio Astuti}
\affil{Bank of Italy, Directorate General for Economics, Statistics and Research}

\date{}

\begin{document}

\maketitle

\begin{abstract}
In this paper we present a technique to couple non-traditional data with statistics based on survey data, in order to partially correct for the bias produced by non-random sample selections. All major social media platforms represent huge samples of the general population, generated by a self-selection process. This implies that they are not representative of the larger public, and there are problems in extrapolating conclusions drawn from these samples to the whole population. We present an algorithm to integrate these massive data with ones coming from traditional sources, with the properties of being less extensive but more reliable. This integration allows to exploit the best of both worlds and reach the detail of typical \rev big data" sources and the representativeness of a carefully designed sample survey.\footnote{The views expressed in this paper are those of the authors and do not involve the responsibility of the Bank of Italy and/or the Eurosystem. We thank  Laura Bartiloro, Costanza Catalano, Marta Crispino, Giovanni D'Alessio, Giuseppe Ilardi, Marco Langiulli, Juri Marcucci, Andrea Neri, Tullia Padellini and Alfonso Rosolia for useful discussions and suggestions.} 
\end{abstract}

\section{Introduction}
In the last two decades we witnessed a revolution in statistics and data-collection processes. The exponential improvement in technological resources and innovations in data analysis allowed us to handle and exploit types of data previously never considered as a useful resource, or not even existing. A prototypical example of new data type is the text produced daily on social media platforms, or on the internet at large. Although the study of language is an old subject and the idea of monitoring media outlets as a proxy for the public opinion is not new either \cite{droba1931methods, mantyla2018evolution}, the volume of information generated online today was simply inconceivable decades ago. The availability of data in such high volumes and detail has made apparent the need of automated methods of analysis, which in turn allowed the further development of the data generation processes. Now \rev big-data" is a research topic on its own, and often it is almost disconnected from traditional statistics. Among the abundance of research papers about the analysis of social media platforms \cite{baylis2015temperature, wu2011says, mittal2012stock, ortigosa2014sentiment, ramage2010characterizing, renault2017}, a very small fraction treats the problem of connecting the conclusions extracted from these new forms of data to the ones derived from traditional statistical studies. An opinion mining study performed on a social network like Twitter can provide extremely detailed information, as we are allowed to hypothetically \rev interview" each individual in the Twitter population; in this regard it is more similar to a census study rather than a sample survey. In addition, it is much less expensive than a traditional statistical survey, which requires careful sample design, questionnaire preparation, and execution by experts. Finally, the greater simplicity allows the study of these independently produced data with a greater frequency, and with less preparation time. 
Even with all these undeniable advantages, however, it should be kept in mind that a social media analysis is not an analysis of the full population, and the same is true for the most common forms of \rev big data". As extensive as the population of users of a social media platform can be, it cannot be as representative of the general population as a carefully designed survey. For example, the population of Twitter users is generated by a self-selection process that introduces evident biases in any conclusion we might draw from a study of the platform \cite{miranda2015twitter}. In addition to not being representative of a more extended population, these self-generated samples do not typically allow for the identification of the single statistical units; while in a survey study the interviewers can easily acquire a host of information about each individual interviewed (and thus the relationships between the variables of interest and any number of auxiliary variables), from \rev big data" samples we can extract only aggregate statistics over a limited number of variable. As an example, we can easily extract the distribution of frequency of appearance of the most used words on a given social media, but it is much less straightforward to study the correlation between these frequencies and the age of the users. 

The introduction of biases during the process of sample creation is a widely studied issue in statistics, although usually in different contexts. One of the most famous examples of a massive sample leading to wrong conclusions pre-dates modern \rev big data" by almost a century: in 1936 the Literary Digest Magazine performed a poll covering more than two million of its readers about presidential election results \cite{squire19881936}. The poll assigned a sweeping victory to the wrong candidate, for two reasons: \textit{i)} the readership of the magazine was not representative of the full American electorate, \textit{ii)} not all the interviewed readers responded to the poll, creating an additional bias. 
A well designed survey about voting intentions performed at the same time would have produced radically different results. Would the availability of such a traditional survey have rendered the Literary Digest Magazine poll useless? This is the question we try to answer here. The findings we report in the following show that the two forms of polling, even though implying opposite outcomes when taken independently, can be integrated to deliver the correct conclusions, and much more detail on the population than what would be possible with the survey alone. More generally, we will show that when we have:
\begin{itemize}
    \vspace{-2mm}
    \item a representative survey revealing, for example, the average of some quantity of interest over a population;
    \vspace{-2mm}
    \item a large-scale, non-representative poll covering an extensive portion of the same population;
    \vspace{-2mm}
    \item some knowledge about the probability of selection of each individual into such a poll,
    \vspace{-3mm}
\end{itemize}
then we can describe in detail the distribution of the quantity of interest over the whole population.
One of the most famous approaches to solve the problem of selection bias is the so called Heckman correction \cite{heckman1979sample, puhani2000heckman}, a technique to partially correct distortions caused by the non-random nature of the observed data. In particular, whenever a correlation is present between a variable of interest and the probability for an individual to be included in the sample, an additional variable is necessary in the model to correct for the bias created. This approach requires modeling the inclusion probabilities as a function of some additional explanatory variables, and in addition some identification condition is necessary in the sample: we have to know both the variables of interest and the variables modeling the inclusion probabilities for any interviewed individual. A slightly different problem is the so-called non-response bias: it is not uncommon that even in a well designed survey, a significant fraction of the sampled individuals choose not to answer to the call to be interviewed. While this non-response does not always imply a bias in the results of the survey, it usually requires a correction treatment similar to the one described above \cite{groves2006nonresponse, groves2008impact}. The careful identification of every statistical unit interviewed, necessary to apply these corrections, is not always possible when using non-traditional data sources like social media.
A good overview of the problem and some alternative solutions are given in \cite{elliott2017inference}.
In this paper we propose a method to compensate for the limited knowledge about the single statistical units exploiting both the amount of data and the details coming from a high-volume source, combining the information extracted from a \rev big data" sample with all prior knowledge about a given population.  
The basic idea is that the demographics of a social media platform, even if composed by self-selected individuals, is often well known (as an example see \cite{social_dem}). It is possible to exploit the external information on the probability to be selected in the sample to partially correct for the bias in the non-representative sample, introducing the sample distribution as a constraint. For example, even though we don't know the age of each Twitter user in a given sample, we could know the probability to be selected in the sample as a function of the age. The distribution of the variables of interest in our sample is equal to the original population distribution \rev projected" through this selection probability.
In \cite{lee2001semiparametric, van2011bayesian} the problem of inference of selection probabilities is discussed, but they mostly limit the analysis to parametric forms of the population distribution and on semi-parametric forms of the unknown selection functions. 
We switch their approach, in that we assume a greater knowledge of the selection function and of the sample distribution, but very little prior knowledge about the population distribution. We employ a non-parametric approach; it is a practicable approach in presence of a \rev big data" sample, due to the considerable detail at which we can resolve the sample distribution and the relative precision with which we know the selection probabilities. 

\section{Theoretical framework}\label{sec:framework}
We assume to have a population with an unknown distribution over a set of variables; from this population each individual is included in a sample with a certain probability, and from the sample we can extract the distribution of a set of variables of interest.  
We divide the statistical variables in two groups $\{C_s\}$ and $\{C_o\}$. The former are the variables influencing the selection of the statistical units in the sample, and could in principle be unobservable. The latter are the variables under study (which can also influence the selection probabilities in the sample). 
The population is described by a distribution $\rho_p(C_o, C_s)$. We are interested in the distribution of the observed variables $\{C_o\}$, obtained as the marginal of $\rho_p(C_o, C_s)$ over the selection variables (for the sake of notation we restrict ourselves to discrete variables):
\begin{equation}
    \underline{\rho_{p}}(C_o) = \sum_{\{C_s\}} \rho_p(C_o, C_s)
\end{equation}
The distribution $\underline{\rho_{p}}(C_o)$ is the unknown we want to estimate, on the basis of prior partial information obtained from an independent source (for example a previous survey study). 
The prior information we need to apply the estimation algorithm amounts to:
\begin{itemize}
    \item some knowledge about the population distribution, e.g. some of its moments; in the following applications we will assume to know the average of the distribution, but higher moments or different information can be considered in the method; 
    \item the probabilities $\rho_s(C_o, C_s)$ to be included in the sample as a function of the observed variables $C_o$ and some potentially unobservable variables $C_s$. These probabilities need to be obtained independently - for example from a demographic of the users in the initial population.
\end{itemize}
With this information we can give an estimation of the probability distribution  $\underline{\rho_{p}}(C_o)$ over the variables of interest in the population, as shown in the following. From the observed sample we have a distribution $\rho_o(C_o)$ which gives a detailed - though not representative - knowledge of the variables $C_o$.\footnote{Such detailed distributions are almost ubiquitous when treating \rev big data" sources, like social media texts, credit cards spending records, or any extensive survey with a self-selection mechanism. } 
The sample distribution $\rho_o(C_o)$ can be derived from the population distribution with a convolution with the selection probabilities $\rho_s(C_o, C_s)$:
\begin{equation}
    \rho_o(C_o) = \sum_{\{C_s\}} \rho_p(C_o, C_s) \rho_{s}(C_o, C_s)
\end{equation}
The last equation can be recast as a constraint on the unknown distribution $\rho_p(C_o, C_s)$, allowing us to exploit the information we have on $\rho_o(C_o)$ and $\rho_s(C_o, C_s)$. Denoting with $P[\bullet ]$ the \rev functional" probability\footnote{The probability to have a given distribution can be defined as the joint probability to have given values for the distribution evalued at every point:  $P[\rho]:= P\left[ \rho(x_1)=\rho_1, \rho(x_2)=\rho_2,...,\rho(x_n)=\rho_n\right]$.} for the distribution $\rho_p$, we can state Bayes' theorem as:
\begin{equation}
\label{bayesprobability}
    P[\rho_p|\rho_o] \propto P[\rho_p] P[\rho_o|\rho_p] 
\end{equation}
where $P[\rho_p]$ is the probability compatible with all the prior knowledge about the distribution $\rho_p$, and $P[\rho_o|\rho_p]$ the probability to observe a distribution $\rho_o$ given a population distribution $\rho_p$. Once we obtain the posterior probability we could in principle calculate the expectation value of every quantity of interest, but these evaluations are usually computationally hard; furthermore the probabilities $P[\rho_p|\rho_o]$ are exponentially concentrated around their maxima, so we can simply derive the best estimate for the probability distribution $\rho_p(C_o, C_s)$ by maximizing the posterior probability \eqref{bayesprobability}. 

We can incorporate all the prior knowledge about the population distribution $\rho_p$ in a maximum entropy prior: assuming for example the prior knowledge to consist of the expectation value of a function  $f(C_o, C_s)$, we obtain the maximum entropy distribution as the solution of the optimization problem \cite{jaynes1957information_1, jaynes1957information_2}:
\begin{equation}
\label{prioroptimization}
    \max_{\{\rho_p\}; \{\lambda_1,\lambda_f\}}\left\{S[\rho_p] + \lambda_1 \left(E_{\rho_p}[1] - 1\right) + \lambda_f  \left(E_{\rho_p}[f] - \overline{f} \right) \right\}
\end{equation}
where $\lambda_1$ and $\lambda_f$ are Lagrange multipliers, $S[\rho_p]$ is the entropy of the distribution $\rho_p$:
\begin{equation}
    S[\rho_p] = \sum_{\{C_o,C_s\}} \rho_p(C_o,C_s) \log\frac{1}{\rho_p(C_o,C_s)}
\end{equation}
and $E_{\rho_p}[\bullet]$ is the expectation value with respect to the probability distribution $\rho_p$:
\begin{equation}
    E_{\rho_p}[1] =  \sum_{\{C_o,C_s\}} \rho_p(C_o,C_s)  \qquad E_{\rho_p}[f] = \sum_{\{C_o,C_s\}} \rho_p(C_o,C_s) f(C_o, C_s)
\end{equation}
Assuming a perfect knowledge\footnote{This is not a strict requirement, in that we could treat this distribution too in a probabilistic way, at the cost of a greater computational burden.} of the selection probabilities $\rho_s$, the likelihood term $P[\rho_o|\rho_p]$ takes the form of a constraint:
\begin{equation}
    P[\rho_o|\rho_p] = \prod_{\{C_o\}} \delta[\rho_o(C_o) - \sum_{\{C_s\}} \rho_p(C_o, C_s) \rho_{s}(C_o, C_s)]
\end{equation}
The meaning of the last equation is simply that $P[\rho_o|\rho_p]$ is null whenever the sample probability distribution $\rho_o(C_o)$ is different from the sum on the right hand side.
Combining this constraint and the optimization problem \eqref{prioroptimization} with another set of Lagrange multipliers, we obtain $\hat{\rho}_p$ as the solution to the maximization problem:
\begin{equation}
    \label{maxent}
     \max_{\{\rho_p\}; \{\lambda_1,\lambda_f, \lambda(C_o)\}}\left\{S[\rho_p] + \lambda_1 \left(E_{\rho_p}[1] - 1\right) + \lambda_f  \left(E_{\rho_p}[f] - \overline{f} \right)  + V[\lambda(C_o), \rho_p] \right\}
\end{equation}
with
\begin{equation}
    V[\lambda(C_o), \rho_p] = \sum_{\{C_o\}} \lambda(C_o)  \left(  \sum_{\{C_s\}} \rho_p(C_o, C_s) \rho_{s}(C_o, C_s) - \rho_o(C_o) \right) 
\end{equation} 
As we can see from equation \eqref{maxent}, we have a number of constraints greater than the number of values in the domain of the observed distribution $\rho_o(C_o)$, thus substantially limiting the form of the estimate $\hat{\rho}_p$. The solution to the problem \eqref{maxent} has the form:
\begin{equation}
    \label{estimate}
    \hat{\rho}_p(C_o, C_s; \hat{\lambda}_f, \hat{\lambda}(C_o)) = N^{-1} \exp{\left[\hat{\lambda}_f f(C_o, C_s) + \hat{\lambda}(C_o) \rho_s(C_o, C_s) \right]}
\end{equation}
where $N$ is a normalization constant:
\begin{equation}
N = \sum_{\{C_o, C_s\}}  \exp{\left[\hat{\lambda}_f f(C_o) + \hat{\lambda}(C_o) \rho_s(C_o, C_s) \right]},
\end{equation}
and $\hat{\lambda}_f$ and $\hat{\lambda}(C_o)$ are solutions to the constraint equations:
\begin{equation}
   \sum_{\{C_o,C_s\}} \hat{\rho}_p(C_o, C_s; \hat{\lambda}_f, \hat{\lambda}(C_o))\, f(C_o, C_s) = \overline{f}
\end{equation}

\begin{equation}
    \sum_{\{C_s\}} \hat{\rho}_p(C_o, C_s; \hat{\lambda}_f, \hat{\lambda}(C_o)) \, \rho_{s}(C_o, C_s) = \rho_o(C_o).
\end{equation}
The solution \eqref{estimate} is the most probable estimation of the real population distribution $\rho_p(C_o, C_s)$ in terms of Bayes' posterior probability, assuming the only prior knowledge we have consists of the expectation value $\overline{f}$ and the distributions $\rho_s(C_o, C_s)$ and $\rho_o(C_o)$.  
As we stated in the beginning we are interested in the distribution of the variables $\{C_o\}$, such that we can restrict our analysis to the marginals:
\begin{equation}
    \underline{\hat{\rho}_p}(C_o; \hat{\lambda}_f, \hat{\lambda}(C_o)) = \sum_{\{C_s\}} \hat{\rho}_p(C_o, C_s; \hat{\lambda}_f, \hat{\lambda}(C_o))
\end{equation}
To summarize the necessary ingredients for the application of the proposed algorithm, the additional prior information is composed of:
\begin{itemize}
    \item some prior about the distribution we want to estimate (typically the mean, but higher moments or other features of the distribution can be taken into consideration);
    \vspace{-2mm}
    \item the probability to be included in the sample as a function of some (possibly unobservable) variables $C_s$;
    \vspace{-2mm}
    \item the sample distribution of the variables of interest $C_o$.
\end{itemize}
The output of the model is the most probable distribution compatible with the information available.
The algorithm exploits the prior information estimating a population distribution which has to satisfy two constraints: the equality of the estimated moments with the ones independently known, and the production of the right sample distribution when the probabilities of selection $\rho_{s}(C_o, C_s)$ are taken into account.
In the following sections we will apply the method described above to a set of simulated samples and to some real distributions extracted from a social media data-set.

\section{Simulated samples}\label{sec:simulation}
To test the method described in Section \ref{sec:framework} we will apply it to a set of simulated population/sample pairs. In each case we will evaluate the information gain with respect to two naive estimations: the first obtained assuming the population to be distributed exactly as the sample (which is what is effectively done when working with big data sources); the second is obtained ignoring the sample distribution, and taking into account only the prior distribution (which is what is effectively done when alternative data sources are ignored). 
The test populations are chosen as Gaussian mixture distributions, obtained as the superposition of 4 normally distributed sub-populations extracted at random with:
\begin{itemize}
    \item means uniformly distributed in the interval $\left[-5,5\right]$;
    \item standard deviations uniformly distributed in the interval $\left[0,1\right]$.
\end{itemize} 
We then constructed a super-population of 2.000 such populations, from each of which we extracted a sample with a second random process. Each individual of the population is selected in the sample with a probability extracted from a uniform distribution in the interval $\left[0 , 1\right]$. This probability is dependent on the Gaussian sub-population the individual is selected from. The result of this procedure is a sample distribution which is effectively a new Gaussian mixture with different weights for every sub-population. 
As an example we show in Figure \ref{fig: example1} an element in this super-population of population-sample pairs: 
\begin{figure}[H]
    \centering
    \includegraphics[width=1\textwidth]{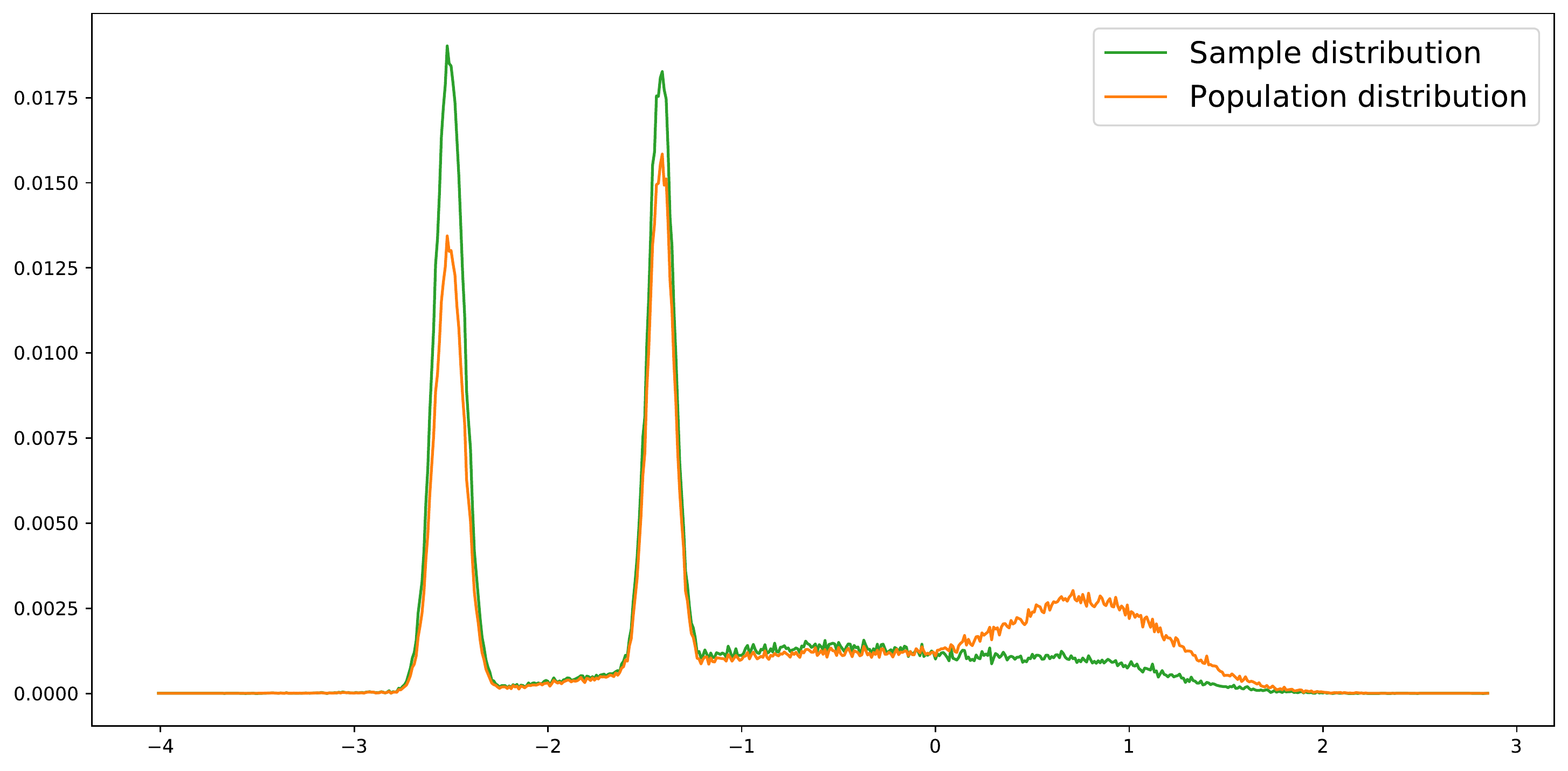}
\caption{Example of a population-sample pair generated by our process.}
\label{fig: example1}
\end{figure}
Once the samples are extracted from the simulated populations we discard all the knowledge of the population distributions apart from:
\begin{itemize}
    \item their average, in order to mimic a real-world situation in which we can see only the sample distribution and a summary statistic of the population obtained from an independent survey;
    \item the selection probabilities with which each sample distribution is generated from the population (these are the random probabilities generated in the process of creation of the population-sample pair, but in any realistic application they have to be obtained from an independent demographic of the sample). 
\end{itemize}
In other words, we keep the information we would otherwise need to get from independent sources, such as the summary statistics (in this case the average) and the selection probabilities. 
The input of our model is thus a set of 2.000 randomly distributed couples $(\rho_{o_i}, \rho_{s_i})$ of sample distributions $\rho_{o_i}$ and selection probabilities $\rho_{s_i}$ and an associated set of known average values (results of \rev simulated" surveys on the random populations). We want to study the performance of our model in inferring each population distribution $\rho_{p_i}$ from these inputs.\footnote{It is unessential for the results of the model, but here we assumed the size of the population to be known (every sub-population being selected with a random weight the size of the sample is also a random variable, thus in principle we don't know the size of the population).}  

The algorithm detailed in Section \ref{sec:framework} allows us to infer the most probable distribution $ \underline{\hat{\rho}_p}(x)$ compatible with all (and only) the constraints given, without additional hypothesis and assumptions. An example of the output is shown in Figure \ref{fig: example2}:
\begin{figure}[H]
    \centering
    \includegraphics[width=1\textwidth]{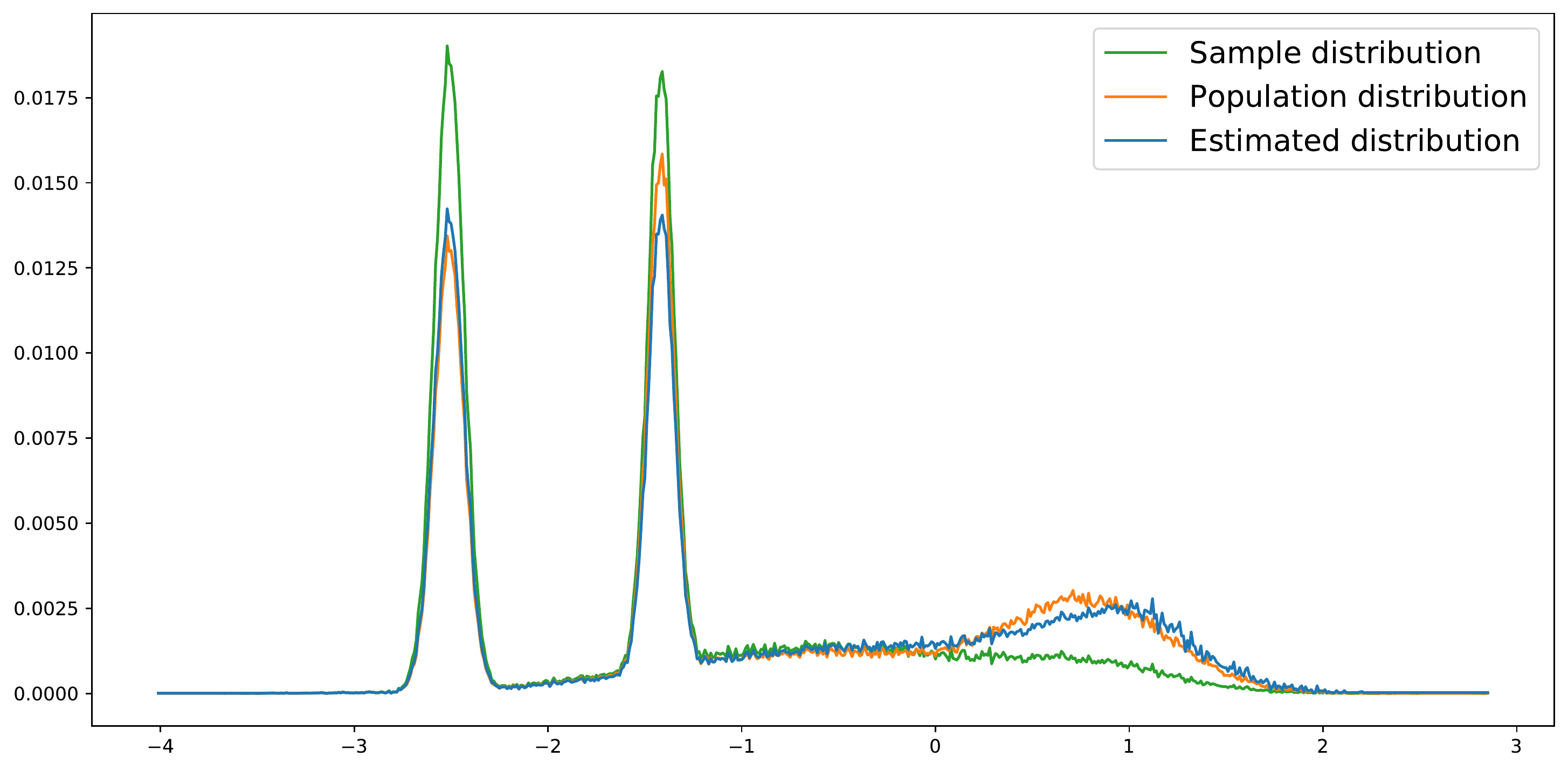}
\caption{Example of prediction for the given population-sample pair.}
\label{fig: example2}
\end{figure}
The availability of a large number of random population-sample pairs allows us to check quantitatively the performance of the model. The measure chosen is the gain in accuracy, i.e. the reduction in the estimation error with respect to a given benchmark estimation. The estimation error is defined as the percentage of individuals who are \rev misplaced" with respect to the correct population distribution; more precisely, given an estimation $\hat{\rho}_p(x)$, we define its error as:
\begin{equation}
    \varepsilon\left(\hat{\rho}_p\right) = \frac{1}{2} \sum_{x} \left| \hat{\rho}_p(x) - \rho_p(x) \right|
\end{equation}
The accuracy gain is evaluated with respect to the estimations made assuming the population to be statistically identical to the sample in the first case, and totally ignoring the sample and considering only the prior knowledge of the population in the second case. 

\begin{figure}[H]
    \centering
    \includegraphics[width=1\textwidth]{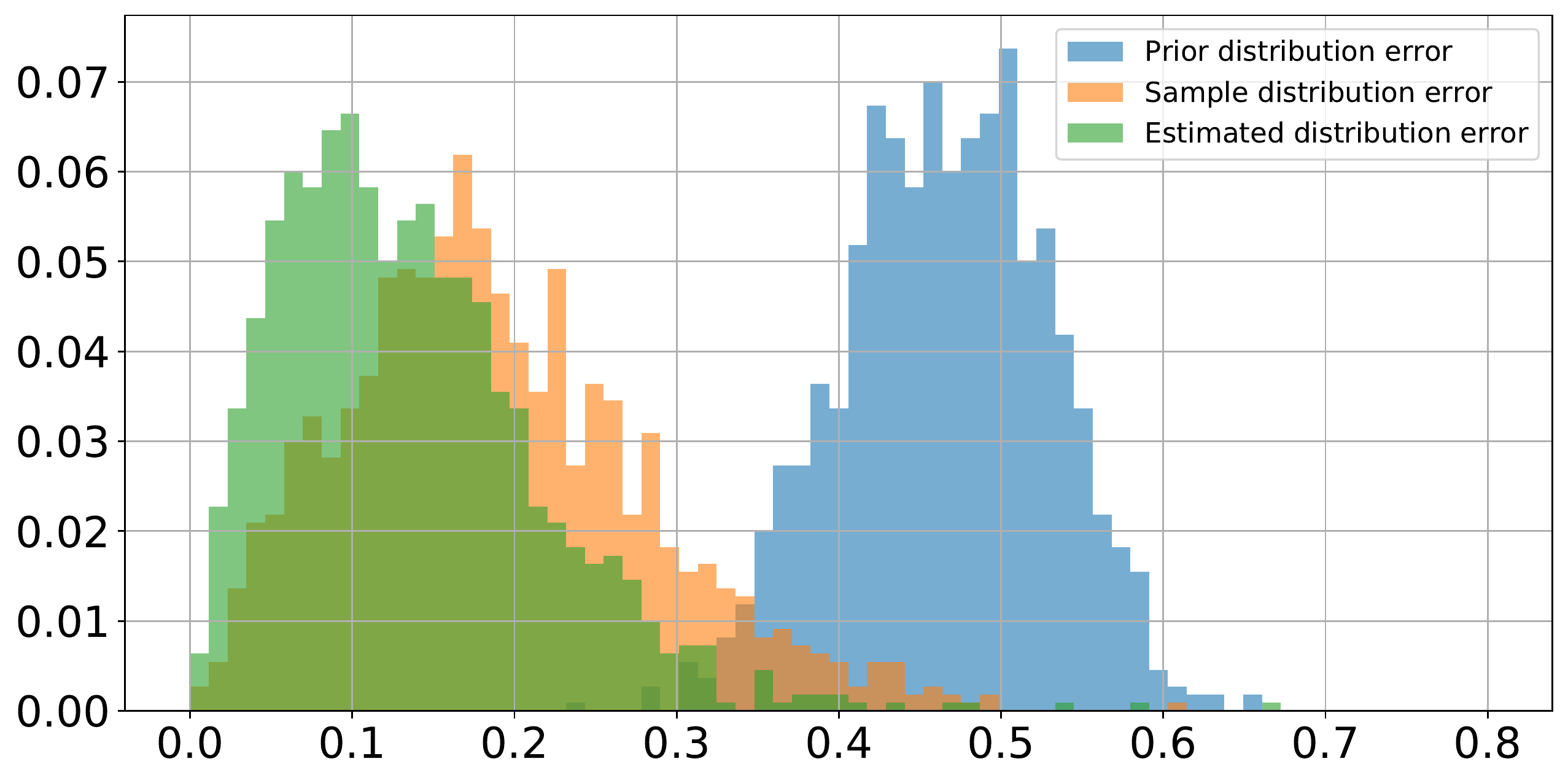}
\caption{Distributions of errors for the simulated distribution/sample pairs.}
\label{fig: errori_sim}
\end{figure}

With respect to the first \rev \emph{population=sample}" benchmark we obtain on average a reduction in the estimation error of 29\% with a best 25-percentile reduction of 52\% and a worst 25-percentile reduction of 10\%. This means that on average we are 29\% more accurate in describing the population than the given benchmark, and in general we have a high likelihood of improving our estimation applying the proposed method as opposed to the naive approach of considering only the sample population. In absolute terms, the error in the benchmark is equal to 19\% on average, while combining the information in the sample and in the prior knowledge of the distribution we get an average error of 13\%.  
The gain in accuracy is bigger when the model is compared with the benchmark consisting only in the prior knowledge of the population distribution. We have an average relative gain in accuracy of 70\%, 84\% and 60\% respectively on average, in the best 25-percentile and in the worst 25-percentile. These high numbers are however mostly explained by the low accuracy of the \rev \emph{prior}" benchmark: considering only the prior knowledge of the population distribution we have an average error of 46\%, with best 25-percentile of 42\% and worst 25-percentile of 51\%.
The result are show in Table \ref{tab:results_benchmark}, where we denoted the two benchmarks \emph{Pure Prior} and \emph{Pure Sample}, while our estimation method is denoted \emph{Prior+Sample}. In addition we show the distribution of errors in Figure \ref{fig: errori_sim}.
\begin{table}[H]
\centering
\begin{tabular}{lcccc}
    \hline
    Model & Mean & 25-th percentile & Median & 75-th percentile \\ 
    \hline
    Pure Prior & 0.46 & 0.42 & 0.46 & 0.51  \\ 
    Pure Sample & 0.19 & 0.12 & 0.18 & 0.24  \\
    Prior+Sample & 0.12 & 0.07 & 0.11 & 0.17  \\
    \hline
\end{tabular}
\caption{Estimation error for the different models} 
\label{tab:results_benchmark}
\end{table}

\section{Twitter samples}\label{sec:twitter}
In this section we want to apply the inference algorithm to a real data-set, consisting of messages extracted from the Twitter platform during the year 2020. For that time period we downloaded all tweets containing at least one element of a list of stopwords, chosen in order to limit the number of tweets without creating bias in their features.\footnote{The stopwords used to restrict the set of tweets are: \emph{\rev dello"}, \emph{\rev fra"}, \emph{\rev lei"}, \emph{\rev nostro"}, \emph{\rev tuo"}, \emph{\rev vostro"}. The choice of stopwords is however not important in our settings, given that the tweets selected with this choice are considered as the population from which samples will be extracted with a different method.}
The main variable considered here - and among the main aspects studied commonly on social media platforms - will be the sentiment expressed in the tweets. The sentiment score is a synthesis of the \rev positivity" or \rev negativity" of a text, designed to represent the emotions expressed in a given document. 
This index has been shown to be useful e.g. in predicting stock market movements, election outcomes and economic trends \cite{mittal2012stock, pagolu2016sentiment, tumasjan2010predicting, angelico2021can}. In its simplest algorithmic implementation \emph{sentiment analysis} consists in assigning a score to each word in a vocabulary (which can be tailored to the subject under analysis), usually in the interval $[-1, 1]$. A low/high score is assigned to words regarded as negative/positive, while all the words considered neutral or not present in the vocabulary will have a null score. The score of any document is then easily computed as the sum of the scores of the component words, normalized with a function of the number of words in the document. In principle assigning a score to every word in a vocabulary (big enough to be useful in a given application) is the most difficult part of the process, but a good number of generic and topic-specific vocabularies are available in the literature, especially for the English language \cite{loughran2011liability, bennani2017home, apel2012information}. We used a vocabulary composed for the Italian language and used in many applications \cite{bruno_voc}. Finally, we normalized the score of each tweet by the number of words in the tweets which were also present in the vocabulary:
\begin{equation}
    S_{t_i} = \frac{ \sum_{w \in (t_i \cap \text{Voc})} S_{w_i}}{\left|t_i \cap \text{Voc} \right|}
\end{equation}
where we denote by $t_i \cap \text{Voc}$ the set of words present in both the tweet $t_i$ and the vocabulary, and with $\left|t_i \cap \text{Voc} \right|$ the cardinality of this set. With this normalization the sentiment scores of the tweets is distributed in the interval $[-1,1]$.

Hence our population consists in a set of tweets sentiment scores, to which we can associate the unique identifier codes of the users who posted them.  
This population substitutes the artificial ones created in Section \ref{sec:simulation}; in particular we don't know the exact form of the probability distribution of the scores\footnote{Although being the score of every tweet the sum of many weakly correlated variables we can expect that the distribution will be approximately Gaussian.}. We perform on this data-set a twofold analysis: first, similarly to what implemented in Section \ref{sec:simulation} we extract a sub-population of users, a large set of random samples and check the performance of the model in inferring the population distribution from the sample ones; secondly we assess the model in an \rev extreme" setting: we try to reconstruct the population distribution from a sample extracted with unknown selection probabilities, with a strong bias on the element selected, and with a large part of the population having a null probability of being selected in the sample.

\subsection{Random samples}
\begin{figure}[H]
    \centering
    \includegraphics[width=0.9\textwidth]{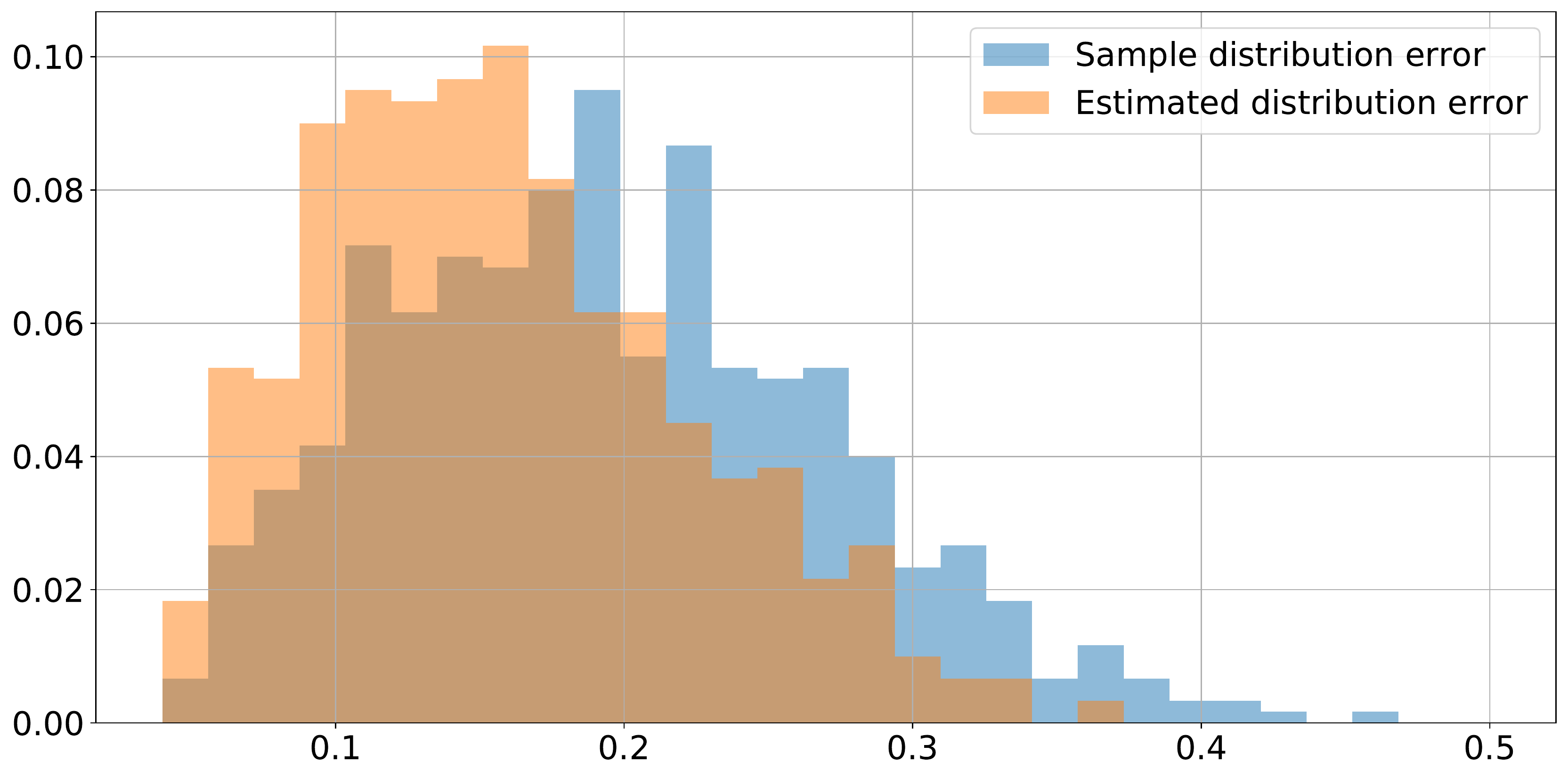}
\caption{Distributions of errors for the simulated Twitter samples.}
\label{fig: errori_twit}
\end{figure}

The first phase of this trial was a selection of a sub-population from the full data-set of tweets. We performed this selection in order to have a population polarized enough that we would obtain heterogeneous samples, while at the same time being able to keep track of the selection probabilities. In the full population, given the sheer number of users taken into consideration we have an inevitable regression toward the mean for almost every significative sample extracted. To avoid this \rev homogenisation" we have to force a bias in the selection by hand, and we will analyze that case in the next section. 
In this section instead we limit the users taken into consideration to the ones with a sufficient divergent set of opinions. First we considered only users with a sufficient number of tweets: we did not consider a user unless he published more than 100 tweets during the period studied. After this first selection we kept the 5 users with the lowest and highest average sentiment scores. This sub-population was composed of more than 6.000 sentiment scores. 
After the selection of the population the analysis proceeded as the one in Section \ref{sec:simulation}: we extracted 600 random samples assigning a random selection probability to each user in the population, and selected tweets in each sample based on these probabilities. Starting from each sample and the knowledge of the associated selection probabilities we used the prior knowledge about the population (here, again, we assume the prior knowledge to consist of the average sentiment score) to adjust the naive estimation based on the sample distribution. The results are similar to the ones obtained for the simulation. Here, given the poor results obtained with the pure prior estimation, we consider only the pure sample benchmark. The sample distributions had an average probability to misassign an individual of 20\%, while the distributions corrected with our method had an average error of 15\%, i.e. a relative improvement of 25\% in the prediction. We show the details of the distributions of errors in Table \ref{tab:results_twitter}, and the full distributions in Figure \ref{fig: errori_twit}. 
\begin{table}[H]
\centering
\begin{tabular}{lcccc}
    \hline
    Model & Mean & 25-th percentile & Median & 75-th percentile \\ 
    \hline
    Pure Sample & 0.20 & 0.14 & 0.19 & 0.25  \\
    Prior+Sample & 0.15 & 0.11 & 0.15 & 0.20  \\
    \hline
\end{tabular}
\caption{Estimation error for the different models} 
\label{tab:results_twitter}
\end{table}

\subsection{Unknown selection probabilities}
In this section we want to expose some of the limits of the model presented: in particular we specified in Section \ref{sec:framework} that to apply this method of inference we need a good knowledge of the selection probabilities upon which the sample is constructed from the population. This is a realistic assumption in the case of social media samples, on which a good demographics is often available, but it is not true in general. As an example we could select a sample of tweets from our population relying only on the date of publication: for example we could select tweets published on weekdays, and ignore the weekends. This will of course introduce a bias, but the choice cannot be modeled without a detailed knowledge of the population. In addition such a sample would give no information about the tweets outside the sample: there could be no tweet at all, they could be identical or completely different from the ones we considered in the sample. 
We can model the situation with an omitted variable $C_s$, upon which the selection in the sample depends in a deterministic way (i.e.: $C_s=1$ means the individual is selected in the sample, $C_s=0$ means it is excluded). The unknown population distribution is $\rho_p(C_o,C_s)$, where $C_o$ is the variable of interest, in this case the sentiment score $S$. 
The selection probabilities are now given by an indicator function $\mathbb{1}(C_s)$. In other words the population distribution can be decomposed as:
\begin{equation}
    \rho_p(C_o,C_s) = W_{C_s}\,  \mathbb{1}(C_s)\, \rho_o(C_o)  + \left(1-W_{C_s}\right) \left[1 - \mathbb{1}(C_s)\right] \rho_{\neg o}(C_o) 
\end{equation}
where $\rho_{\neg o}(C_o)$ is the unobserved distribution of individuals not represented in the sample, and $W_{C_s}\in[0,1]$ is the (unknown) weight of the sample with respect to the total population. In this setting the inference model has to produce a guess about $W_{C_s}$ and $\rho_{\neg o}(C_o)$ with basically no data, so that the final estimation will contain scarce information in addition to the prior one we have about the population. The unobserved distribution $\rho_{\neg o}(C_o)$ will be the maximum entropy distribution compatible with the sample distribution and the constraints given by the prior knowledge about the population. 

Here to create a strong bias we selected a sample including all (and only) tweets with a negative sentiment score, while the population contains all tweets. In this case $C_s=1 - \Theta(C_o)$, the sample is perfectly representative of the population with negative sentiment, but it does not contain any information about positive sentiment tweets. Such a selection introduces a strong bias in the sample, given that the average sentiment score in the population is close to zero. As expected, the naive estimation made assuming the population distribution to be identical to the sample one is very far from accurate: it produces an error rate of 61\%. Trying to correct the estimation with the prior knowledge of the population distribution using as before only the average sentiment of the population gives very modest improvements: we reach an error rate of 58\%. If however we take into account both the population average and standard deviation, we obtain a significant improvement: we find an inferred distribution involving an error rate of 28\%. The seemingly low value is given by the fact that in the portion of population with negative sentiment we can describe the distribution perfectly due to the representative sample; in the portion of population with positive sentiment, on the other hand, the estimation is different from zero and of the right magnitude in order to replicate the correct average, thus performing better than the naively estimated distribution. 
 
\section{Comparison with survey methods}\label{sec:comparison}
In section \ref{sec:framework} we explained how the proposed method allows to estimate the population distribution from some prior knowledge and the sample distribution. In this section we want to focus on the main differences between our approach and the ones used in the context of statistical surveys. In particular, we want to emphasize the different kind of difficulties to overcome in the case of statistical surveys with respect to the case here under study. 
In the former situation we have a much narrower sample, which is however under the strict control of the survey designers. We can, in particular, identify every individual interviewed in the sample, and exploit the functional relations between the variables of interest $C_o$ and any additional variable $C_s$ to correct the selection bias. An example of selection bias is given by non-response or under reporting by some individuals in the sample. To correct for this problem in a survey, we can estimate the probability to receive a valid answer as a function  of the auxiliary variables $C_s$. 
Once this probability has been estimated, to correct for non-response bias we can assign more weight to the answers given by individuals with a low probability to respond, by dividing the designed survey weights by the probability value itself \cite{d2015income}. In this way we can impute the missing responses with the ones given by individuals assumed to be statistically similar.
Another common technique to correct for selection bias in statistical surveys is the so-called \emph{calibration}, which consists in changing the statistical weights assigned to each class of auxiliary variables in order to reproduce known distributions of these variables in the population under study, or some known population totals \cite{d2015income, deville1992calibration}. As an example, knowing the gender of the interviewed individuals and the fact that the population under study is gender-balanced, a gender-unbalanced sample can be adjusted changing the weights such as to over-sample the less represented genders. When population totals are taken into consideration, the principle is similar to the one proposed in section \ref{sec:framework}, but also in this case there are important differences. The method is based on corrections of the sample weights, which are also in this case expressed as functions of a set of auxiliary variables $C_s$. The correct weights are chosen in order to minimize a given distance from the original ones, in the space of all possible weights satisfying the constraints given by the known population totals.
Again, this requires the identification of the single statistical units in the sample. This step is necessary in order to construct the statistical weights used to balance the sample. 
In addition, the possibility to correct the results of the survey by changing the statistical weights is based on the fact that we know the value of the variables of interest $C_o$ and the auxiliary variables $C_s$ for each individual in the sample, thus we are able to exploit the correlation between the two sets of variables to correct the bias in $C_o$ from the knowledge of $C_s$. 

For alternative data sources the identification of the single statistical units is not always possible. The difficulty in carefully tracking every individual in the sample implies also the impossibility to make a direct connection between auxiliary variables and variables of interest. 
In general, even when information about the auxiliary variables $C_s$ is present, it takes the form of a sample distribution independent from the distribution of the variables of interest $C_o$. As such, we cannot conclude anything about the correlation between the two sets of variables in the sample. This in turn implies that models commonly used for statistical surveys, such as the ones described above, are not applicable in the more general situations where the method detailed in \ref{sec:framework} is available. 
The approach proposed here exploits the knowledge of the probability to be selected in the sample as a function of the auxiliary variables $C_s$, independently of their relation with the variables $C_o$. This probability of selection contains information about the bias of the sample with respect to the general population, and in turn this information can be combined with the prior knowledge of the population distribution to (probabilistically) correct for the bias.

\section{Conclusions}\label{sec:concl}
In this paper we proposed an algorithm to combine non-traditional, \rev big data" sources with more conventional statistics, coming from survey studies. While information coming from non-traditional data sources like social media is almost inevitably biased with respect to the total population, it provides a high level of detail on the distribution of any variable of interest on the sub-population sampled. On the other hand, survey data are much more reliable, but they are usually focused on few observables, like the average of the distribution of the variables of interest over the population. We can use this unbiased information to \rev tilt" the distribution obtained from non-traditional (potentially biased) data, in order to make an estimation of the probability distribution which would be obtained with a detailed knowledge of the full population. 
We applied this algorithm to a set of simulated populations, and to a real one extracted from Twitter data. In both cases we created a population of samples extracted with random probabilities, to evaluate the performance of our model with respect to an estimation based only on sample data. In both cases we found estimated distributions much closer to the real ones, with an average relative reduction of the error rates of 29 percentage points for the simulated populations and of 25 percentage points for the Twitter population. The algorithm can be applied to translate results from non-traditional data sources to the general population, reducing the bias inherent in these data sources, or to combine results from survey-based analyses with additional information coming e.g from social media.

\printbibliography 
 
 \clearpage
 \appendix

\end{document}